%% file: crab_calibration.tex
\documentclass[twocolumn,longauth, traditabstract]{aa}

\input Planck.tex

\usepackage[nonamebreak]{natbib}
\usepackage[stable]{footmisc}
\usepackage{amsmath}
\usepackage{txfonts}
\usepackage{placeins}
\bibpunct{(}{)}{;}{a}{}{,}
\usepackage{graphicx}
\usepackage{epstopdf}
\usepackage{ifthen} 
\usepackage[breaklinks, colorlinks, citecolor=blue]{hyperref}
\usepackage{overpic}
\usepackage{fixltx2e}
\usepackage{rotating}
\usepackage{subfigure}
\usepackage{epsfig}
\usepackage{multirow}
\usepackage{array}
\usepackage{color}
\usepackage[usernames,dvipsnames]{xcolor}
\usepackage{psfrag}
\usepackage[T1]{fontenc}
\usepackage{url}
\usepackage{breakurl}

\usepackage{xcolor} 
\usepackage{makecell} %

\newcolumntype{L}[1]{>{\raggedright\let\newline\\\arraybackslash\hspace{0pt}}m{#1}}
\newcolumntype{C}[1]{>{\centering\let\newline\\\arraybackslash\hspace{0pt}}m{#1}}
\newcolumntype{R}[1]{>{\raggedleft\let\newline\\\arraybackslash\hspace{0pt}}m{#1}}

\newcommand{\expe}[1]{{\sc{#1}}}

\def\psigal{\ifmmode{\psi_{\rm Gal}}\else{$\psi_{\rm Gal}$}\fi}
\def\avgpsigal{\ifmmode{\overline{\psi}_{\rm Gal}}\else{$\overline{\psi}_{\rm Gal}$}\fi}
\def\deltapsigal{\ifmmode{\Delta\psi_{\rm Gal}}\else{$\Delta\psi_{\rm Gal}$}\fi}
\def\deltapsigalstat{\ifmmode{\Delta\psi_{\rm Gal}^{\rm stat.}}\else{$\Delta\psi_{\rm Gal}^{\rm stat.}$}\fi}
\def\deltapsigalsys{\ifmmode{\Delta\psi_{\rm Gal}^{\rm sys.}}\else{$\Delta\psi_{\rm Gal}^{\rm sys.}$}\fi}
\def\meanangle{\ifmmode{-88.19}\else{$-88.19$}\fi}
\def\sigmangle{\ifmmode{0.33}\else{$0.33$}\fi}
\def\errmax{3.89}\def\errmaxarcmin{233.5}
\def\errstd{1.24}\def\errstdarcmin{74.6}
\def\errgrd{0.33}\def\errgrdarcmin{20.1}
\def\errceb{0.28}\def\errcebarcmin{16.8}
\def\errctb{0.23}\def\errctbarcmin{13.8}
\def\errtbn{0.12}\def\errtbnarcmin{7.2}

\definecolor{mywarningcolor}{RGB}{208,59,32} 
\newcommand{\warning}[1]{} 

\begin{document}

\topmargin=-1cm
\oddsidemargin=-1cm
\evensidemargin=-1cm
\textwidth=17cm
\textheight=25cm
\raggedbottom
\sloppy

\definecolor{Blue}{rgb}{0.,0.,1.}
\definecolor{LightSkyBlue}{rgb}{0.691,0.827,1.}
\definecolor{Red}{rgb}{1.,0.,0.}
\definecolor{Green}{rgb}{0.,1.,0.}
\definecolor{Try}{rgb}{0.15,0.,1}
\definecolor{Black}{rgb}{0., 0., 0.}

\title{Absolute calibration of the polarisation angle for future CMB $B$-mode experiments from current and future measurements of the Crab nebula}

\author{J.~Aumont\inst{\ref{IRAP}}
\and J.~F.~Mac\'{i}as-P\'erez\inst{\ref{LPSC}}
\and A.~Ritacco\inst{\ref{IRAM}}
\and N.~Ponthieu \inst{\ref{IPAG}}
\and A.~Mangilli \inst{\ref{IRAP}}}

\institute{
IRAP, Universit\'e de Toulouse, CNRS, CNES, UPS, (Toulouse), France\label{IRAP}
\and
Laboratoire de Physique Subatomique et de Cosmologie, Universit\'e Grenoble Alpes, CNRS, 53, av. des Martyrs, Grenoble, France\label{LPSC}
\and
Institut de RadioAstronomie Millim\'etrique (IRAM), Granada, Spain \label{IRAM}
\and
Univ. Grenoble Alpes, CNRS, IPAG, 38000 Grenoble, France\label{IPAG}
}

\abstract{A tremendous international effort is currently dedicated to observing the so-called $B$-modes of the Cosmic Microwave Background (CMB) polarisation. If measured, this faint signal imprinted by the primordial gravitational wave background, would be the smoking-gun of the inflation epoch and also quantify its energy scale, providing a rigorous test of fundamental physics far beyond the reach of accelerators. At the unprecedented sensitivity level that the new generation of CMB experiments aims to reach, every uncontrolled instrumental systematic effect will potentially result in an analysis bias that is larger than the much sought-after CMB $B$-mode signal. The absolute calibration of the polarisation angle is particularly important in this sense, as any associated error will end up in a leakage from the much larger $E$ modes into $B$ modes. The Crab nebula (Tau A), with its bright microwave synchrotron emission, is one of the few objects in the sky that can be used as absolute polarisation calibrators. In this paper we review the best current constraints on its polarisation angle from 23 to 353\,GHz, at typical angular scales for CMB observations, from WMAP, XPOL, Planck and NIKA data. These polarisation angle measurements are compatible with a constant angle of $\meanangle{}\,^\circ\pm\sigmangle{}\,^\circ$. We study the uncertainty on this mean angle, making different considerations on how to combine the individual measurement errors. For each of the cases, we study the potential impact on the CMB $B$-mode spectrum and on the recovered $r$ parameter, through a likelihood analysis. We find that current constraints on the Crab polarisation angle, assuming it is constant through microwave frequencies, allow to calibrate experiments with an accuracy enabling the measurement of $r\sim0.01$. On the other hand, even under the most optimistic assumptions, current constraints will lead to an important limitation for the detection of $r\sim10^{-3}$. New realistic measurement of the Crab nebula can change this situation, by strengthening the assumption of the consistency across microwave frequencies and reducing the combined error.
}


\authorrunning{}
\titlerunning{Calibration of the angle for CMB experiments from measurements of the Crab nebula}
\maketitle


\section{Introduction}

The polarisation of the Cosmic Microwave Background (CMB) anisotropies offers a
powerful way to investigate the early Universe. In particular, the so called
primordial CMB polarisation $B$-modes can only be generated by 
primordial gravitational waves (tensor perturbations)
\citep{polnarev1985polarization, 1997PhRvL..78.2054S} arising from
an early inflationary epoch \citep{PhysRevD.23.347, linde1982new}.  
Therefore, the detection of the primordial CMB $B$-modes would constitute a direct proof of inflation, opening a window to new physics. 
However, they are expected to be much fainter (more than an order of magnitude, hence much more difficult to detect) 
than the CMB $E$-modes polarisation anisotropies that are produced by scalar (density) perturbations 
at recombination \citep{hu1997,hu2002}. The CMB polarisation $E$-modes
have been accurately measured by the Planck satellite \citep{2016A&A...594A..11P} and their spectrum is about a factor of 100 fainter than the power spectrum of the CMB temperature anisotropies \citep{2016A&A...594A..11P}.

In the last decade the quest for the CMB polarisation $B$-modes
has become one of the major aims of observational cosmology, leading
to very active instrumental developments and to a large number of CMB
experiments \citep[e.g.][]{bicep2,polarbear,sptpol,actpol}.
The goal of these experiments is to measure the tensor-to-scalar ratio $r$, 
given by the relative amplitude of the primordial tensor and scalar perturbations,
that is directly related to the energy scale of inflation.
Lately, \citet{bicepplanck2015,bicep2016} set a 95\% upper limit for the
detection of the tensor to scalar ratio of $r<0.07$.

Future CMB experiments aiming at measuring the primordial $B$-modes target $r$ values ranging from $10^{-2}$ to $10^{-4}$ \citep[e.g.][]{qubic,advancedact,quijote,bicep3,simonsarray,spt3g,class,piper,spider2,cmbs4,core,litebird}.
Although great efforts are made to reach such
low signal by constantly improving instrumental sensitivity,
residual foreground emission and instrumental systematic effects might limit
the final results. The former has been widely discussed in the literature
\citep[see][and references therein]{2007PhRvD..75h3508A,2009A&A...503..691B,2016JCAP...03..052E}. 

In terms of instrumental systematic effects 
one of the main challenges for future ground, balloon and satellite CMB polarisation experiments
is the accurate calibration of the absolute polarisation angle. The most common strategy to accurately tackle these calibration errors in CMB experiments is the minimisation of the $C_\ell^{TB}$ and $C_\ell^{EB}$ spectra, for which no cosmological signal is expected from standard cosmology parity-invariant physical processes. This strategy has two main limitations: (1) Galactic foregrounds \citep{pipxxx} and uncontrolled systematics can produce non-zero $TB$ and $EB$ spectra and (2) non-standard cosmological mechanisms can produce non vanishing $C_\ell^{TB}$ and $C_\ell^{EB}$ \citep[as for example cosmic birefringence, chiral gravity, see e.g.][]{parityviolation,pipxlix}, that next generation CMB experiments would like to characterise. 

In this context, it might be interesting to use an \emph{external} calibration source for the absolute polarisation angle. This calibration could thus be achieved using observations of well known polarised sources like the Crab nebula (Tau A) \citep{2016IJMPD..2540008K}, which is the brightest polarised
astrophysical object in the microwave sky at angular scales of
few arcminutes.

The Crab Nebula is a plerion-type supernova remnant emitting a highly polarised synchrotron signal \citep{1978A&A....70..419W,1991ApJ...368..463M}
from radio to millimeter wavelengths \citep{macias2010}.
A recent study by \citet{crabnika} has demonstrated that the Crab nebula synchrotron emission from radio to millimeter wavelengths is well characterised by a single power law both in temperature and polarisation, which would indicate that a single population of relativistic electrons is 
responsible for the emission of the nebula. As a consequence the degree and angle of polarisation of the Crab nebula are expected to be constant across frequencies in this range, making the Crab nebula a potential polarisation standard.

In this paper we study in details the current constraints on the Crab polarisation angle and we discuss how they can be used to perform an absolute calibration of the polarisation angle of CMB experiments. We then derive the expected systematic uncertainties on the measured tensor-to-scalar ratio $r$. The paper is organised as follows: in Sect.~\ref{sect:measurements} we review the current best constraints on the Crab nebula microwave polarisation angle from 23 to 353\, GHz. In Sect.~\ref{sec:combined}, we discuss several cases corresponding to different assumptions that can be made on those measurement uncertainties, in order to get the combined error on the Crab nebula polarisation angle. We derive in Sect.~\ref{sec:ebmixing} the spurious CMB $B$-mode signal coming from $E$ to $B$ mixing, if the Crab nebula was to be used as a calibrator for the absolute polarisation angle with such uncertainties. Sect.~\ref{sec:likelihood} presents a likelihood analysis in order to express the mis-calibration errors in terms of biases on the measurement of the tensor to scalar ratio $r$, and we finally discuss our conclusions in Sect.~\ref{sec:conclusion}.


\begin{figure}
\centering
\includegraphics[width=\columnwidth]{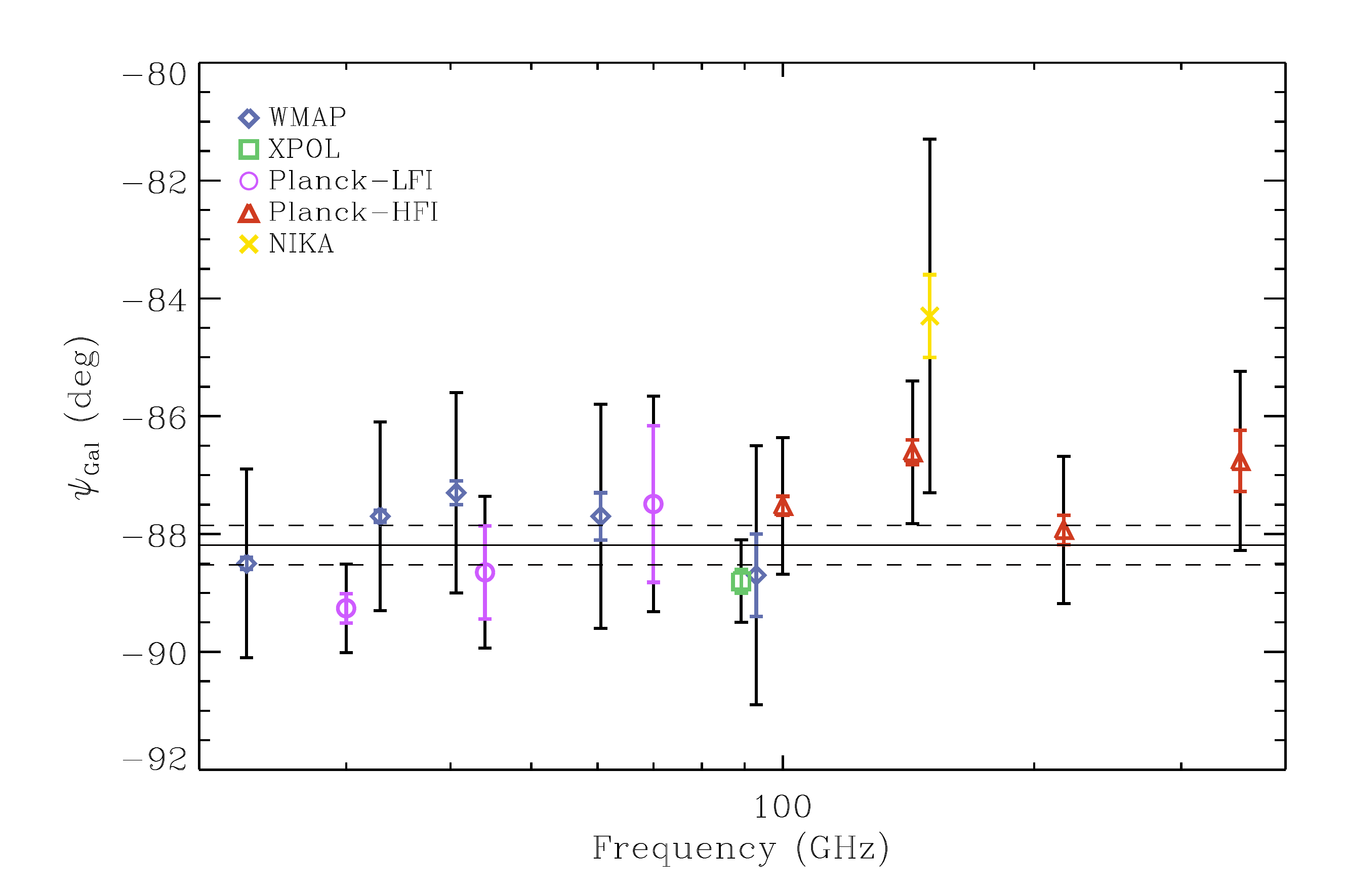}
\caption{\footnotesize Measurements of the Crab nebula polarisation angles from Table~\ref{tab:measures} for \expe{Wmap} (blue diamonds), \expe{Xpol} (green square), \expe{Planck-LFI} (purple circles), \expe{Planck-HFI} (red triangles) and \expe{Nika} (yellow crosses). Statistical error bars \deltapsigalstat{} are colored and the total error budget including systematics ($\deltapsigalstat+\deltapsigalsys$, ground systematics for \expe{Planck-HFI}, corresponding to the \texttt{ground} case of Sect.~\ref{sec:combined}) are in black. The solid and dashed black horizontal lines indicate the weighted mean polarisation angle and its $\pm1\,\sigma$ uncertainty, $\overline{\psi}_{\rm Gal}=\meanangle{}\,^\circ\pm\sigmangle{}\,^\circ$. \label{fig:angles}}
\end{figure}

\section{Crab polarisation angle measurements}
\label{sect:measurements}

\citet{crabnika} gives a compendium of the Crab nebula polarisation angle measurement in Galactic coordinates \psigal{}, from 23 to 353\,GHz. It introduces the \expe{Nika} measurement at 150\,GHz and recomputes the \expe{Planck-HFI} angles (100, 143, 217 and 353\,GHz) in a improved analysis with respect to \citet{crablfi}, based on the \expe{Planck} 2018 maps \citep{planck2016-l03}. \citet{crabnika} also includes measurements by \expe{Wmap} \citep[23, 33, 44, 61 and 94\,GHz,][]{crabwmap}, \expe{Xpol} \citep[90\,GHz,][]{crabxpol} and \expe{Planck-LFI} \citep{crablfi}. In the following, we have chosen not to take into account \expe{Polka} \citep{polka} data point presented in \citep{crabnika}, which is a clear outlier.

The \psigal{} values presented in \citet{crabnika} are reported in Table~\ref{tab:measures}, together with their associated statistical and systematic uncertainties. For \expe{Planck-HFI}, we refer to the pre-flight errors on the absolute calibration of the polarisation angle \citep{rosset} as the \emph{ground} calibration error. These absolute calibration errors were later refined at 100, 143 and 217\,GHz in \citet{pipxlvi} using $C_\ell^{TB}$ and $C_\ell^{EB}$ minimisation, for which no cosmological signal is expected in the abscence of parity violating processes (although Galactic signals \emph{could} produce a non-zero $C_\ell^{TB}$ or $C_\ell^{EB}$ signal, \citep{pipxxx}). We refer to these errors as $TB$ and $EB$, respectively. No $TB$ and $EB$ error were assessed for the 353\,GHz channel, so that we will always assign this channel measurement with the \expe{Planck-HFI} ground uncertainty.

The Crab polarisation angle values in Table~\ref{tab:measures} are compatible with a constant angle from 23 to 353\,GHz (Fig.~\ref{fig:angles}), computed as the inverse-noise weighted average considering the \emph{ground} systematic uncertainties: 

\begin{equation}
\avgpsigal=\frac{\sum_i\psi_{\rm Gal}^i/(\Delta\psi_{\rm Gal}^i)^2}{\sum_i1/(\Delta\psi_{\rm Gal}^i)^2}\pm\sqrt{\frac{1}{\sum_i1/(\Delta\psi_{\rm Gal}^i)^2}}\nonumber
\end{equation}
\begin{equation}
=\meanangle{}\,^\circ\pm\sigmangle{}\,^\circ,
\end{equation}

 \noindent where $\psi_{\rm Gal}^i$ and $\Delta\psi_{\rm Gal}^i$ are the individual measurements and their errors presented in Table~\ref{tab:measures}. The \avgpsigal{} value differs slightly from the one reported in \citet{crabnika}, as we excluded the outlying \expe{Polka} measurement from the present analysis. 

 To derive this \avgpsigal{} value, we consider that for each individual measurement the total error \deltapsigal{} is the sum of the statistical error \deltapsigalstat{} and the systematic error \deltapsigalsys{}. \warning{I did the linear sum of the statistical and systematic uncertainties, not the quadratic one. Is that really what we want to do?}  

\begin{table*}
\begin{center}\setcellgapes{1pt}\makegapedcells
\begin{tabular}{cccccccc}
\hline
\hline
\multirow{2}{*}{Experiment} & \multirow{2}{*}{$\nu$ (GHz)} & Beam & \multirow{2}{*}{\psigal (deg)} & Statistical & \multicolumn{3}{c}{Systematic \deltapsigalsys (deg)} \\
\cline{6-8}
 &  & size &  & \deltapsigalstat (deg) & Ground & $EB$ & $TB$ \\
\hline
\multirow{5}{*}{\expe{Wmap}}       & 23  & $53'$  & $-88.5$  & 0.1  & 1.5  & $-$  & $-$  \\
 							       & 33  & $40'$  & $-87.7$  & 0.1  & 1.5  & $-$  & $-$  \\
 							       & 41  & $31'$  & $-87.3$  & 0.2  & 1.5  & $-$  & $-$  \\
 							       & 61  & $21'$  & $-87.7$  & 0.4  & 1.5  & $-$  & $-$  \\
 							       & 94  & $13'$  & $-88.7$  & 0.7  & 1.5  & $-$  & $-$  \\
\hline
\expe{Xpol}                        & 90  & $27''$ & \ $-88.8^\star$ & 0.2 & 0.5 & $-$ & $-$      \\
\hline
\multirow{3}{*}{\expe{Planck-LFI}} & 30  & $33'$  & $-89.26$ & 0.25 & 0.5  & $-$ & $-$ \\
								   & 44  & $27'$  & $-88.65$ & 0.79 & 0.5  & $-$ & $-$ \\
								   & 70  & $13'$  & $-87.49$ & 1.33 & 0.5  & $-$ & $-$ \\
\hline
\multirow{4}{*}{\expe{Planck-HFI}} & 100 & $10'$  & $-87.52$ & 0.16 & 1.00 & 0.63 & 0.22 \\
								   & 143 & $7'$   & $-86.61$ & 0.21 & 1.00 & 0.42 & 0.27 \\
								   & 217 & $5'$   & $-87.93$ & 0.25 & 1.00 & 0.51 & 0.83 \\
								   & 353 & $5'$   & $-86.76$ & 0.52 & 1.00 & $-$  & $-$  \\
\hline
\expe{Nika}                        & 150 & $18''$ & \ $-84.3^\bullet$ & 0.7 & 2.3 & $-$      & $-$      \\
\hline
\hline
\multicolumn{2}{l}{\scriptsize $^\star$ Convolved with a $10'$ Gaussian} & & & & & &
\\[-5pt]
\multicolumn{2}{l}{\scriptsize $^\bullet$ Computed with aperture photometry techinques within $9'$}& & & & & &

\end{tabular}
\caption{\footnotesize Compendium of the submillimetre Crab nebula polarisation angle measurements in Galactic coordinates for \expe{Wmap} \citep{crabwmap}, \expe{Xpol} \citep{crabxpol}, \expe{Planck-LFI} \citep{crablfi} and \expe{Planck-HFI} and \expe{Nika} \citep{crabnika}. In the case of \expe{Planck-HFI}, the so-called ground systematic uncertainties come from \citet{rosset}. The systematic uncertainties named $EB$ and $TB$ are derived from the $C_\ell^{EB}$ and $C_\ell^{TB}$ minimisation presented in~\citet{pipxlvi}.\label{tab:measures}}
\end{center}
\end{table*}


\begin{table}
\begin{center}\setcellgapes{1pt}\makegapedcells
\begin{tabular}{ccc}
\hline
\hline
Case & \deltapsigal (deg) & \deltapsigal(arcmin)\\
\hline
\texttt{max}       & $\errmax{}$ & $\errmaxarcmin{}$ \\
\texttt{stddev}    & $\errstd{}$ & $\errstdarcmin{}$ \\
\texttt{ground}    & $\errgrd{}$ & $\errgrdarcmin{}$ \\
\texttt{EB}        & $\errceb{}$ & $\errcebarcmin{}$ \\
\texttt{TB}        & $\errctb{}$ & $\errctbarcmin{}$ \\
\texttt{TB+future} & $\errtbn{}$ & $\errtbnarcmin{}$ \\
\hline
\hline
\end{tabular}
\caption{\footnotesize Summary of the combined errors \deltapsigal{} on the Crab polarisation angle \avgpsigal{} for the different cases presented in Sect.~\ref{sec:combined}.\label{tab:combined}}
\end{center}
\end{table}

\section{Combined uncertainty on the Crab polarisation angle}
\label{sec:combined}

In order to use the Crab nebula submillimetre polarisation angle \avgpsigal{} as an absolute angle calibrator for CMB measurements, we are interested in the constraints on its uncertainty \deltapsigal{}, assessed from the measurements presented in Sect.~\ref{sect:measurements}. Given the relatively small number of measurements and the variety of instruments, observing conditions and data processing, there is no unique way to combine them all into a single result with a well defined uncertainty. We therefore propose and test several combinations of these measurements to assess the combined uncertainty \deltapsigal{}: \\[-8pt] 

\noindent {\tiny \textbullet}\ \texttt{max}: we do not assume that the Crab polarisation angle \psigal{} is constant from 23 to 353\,GHz and we take the combined error \deltapsigal{} as the maximum difference between the inverse-noise weighted mean \avgpsigal{} and an individual measurement (the \expe{Nika} measurement). The combined error is in this \texttt{max} case $\deltapsigal=\errmax{}\,^\circ$ ($\errmaxarcmin{}$\,arcmin)\\[-8pt] 

\noindent {\tiny \textbullet}\ \texttt{stddev}: we do not assume that the Crab polarisation angle \psigal{} is constant from 23 to 353\,GHz. We assume that the error on its value is dominated by the inter-frequency variations. We thus take the standard deviation among the individual measurements to be the combined error on the Crab polarisation angle. In this \texttt{stddev} case, the combined error is $\deltapsigal=\errstd{}\,^\circ$ ($\errstdarcmin{}$\,arcmin)\\[-8pt] 

\noindent {\tiny \textbullet}\ \texttt{ground}: we assume that the Crab polarisation angle \psigal{} is constant between 23 and 353\,GHz. The combined error is thus taken as the error on the inverse-noise weighted mean. In the \texttt{ground} case, we take the pre-flight assessment of the error on the absolute calibration angle \citep{rosset} as being the dominant systematic error \deltapsigalsys{} for \expe{Planck-HFI}. The combined error is in this case $\deltapsigal=\errgrd{}\,^\circ$ ($\errgrdarcmin{}$\,arcmin)\\[-8pt] 

\noindent {\tiny \textbullet}\ \texttt{EB}: as for the \texttt{ground} case, the Crab polarisation angle is assumed constant. The difference with the \texttt{ground} case is that we use the $C_\ell^{EB}$ minimisation assessment of the error \deltapsigalsys{} for the 100, 143 and 217\, GHz \expe{Planck-HFI} channels \citep{pipxlvi}. For the other experiments and for the \expe{Planck-HFI} 353\,GHz channel, the \texttt{ground} errors are used. The resulting combined error is $\deltapsigal=\errceb{}\,^\circ$ ($\errcebarcmin{}$\,arcmin)\\[-8pt] 

\noindent {\tiny \textbullet}\ \texttt{TB}: same as \texttt{EB}, but with the $C_\ell^{TB}$ minimisation \deltapsigalsys{} \citep{pipxlvi}. The resulting combined error is $\deltapsigal=\errctb{}\,^\circ$ ($\errctbarcmin{}$\,arcmin)\\[-8pt] 

\noindent {\tiny \textbullet}\ \texttt{TB+future} same as \texttt{TB} but adding 2 future measurements points having each a total error of $\deltapsigalstat+\deltapsigalsys=0.2\,^\circ$. The combined error, assuming a constant polarisation angle for the Crab is in this case $\deltapsigal=\errtbn{}\,^\circ$ ($\errtbnarcmin{}$\,arcmin)\\[-8pt]

The values of the combined error \deltapsigal{} on the Crab polarisation angle \avgpsigal{} for the different cases presented above are summarised in Table~\ref{tab:combined}.


\begin{figure*}
\centering
\includegraphics[height=0.33\textwidth]{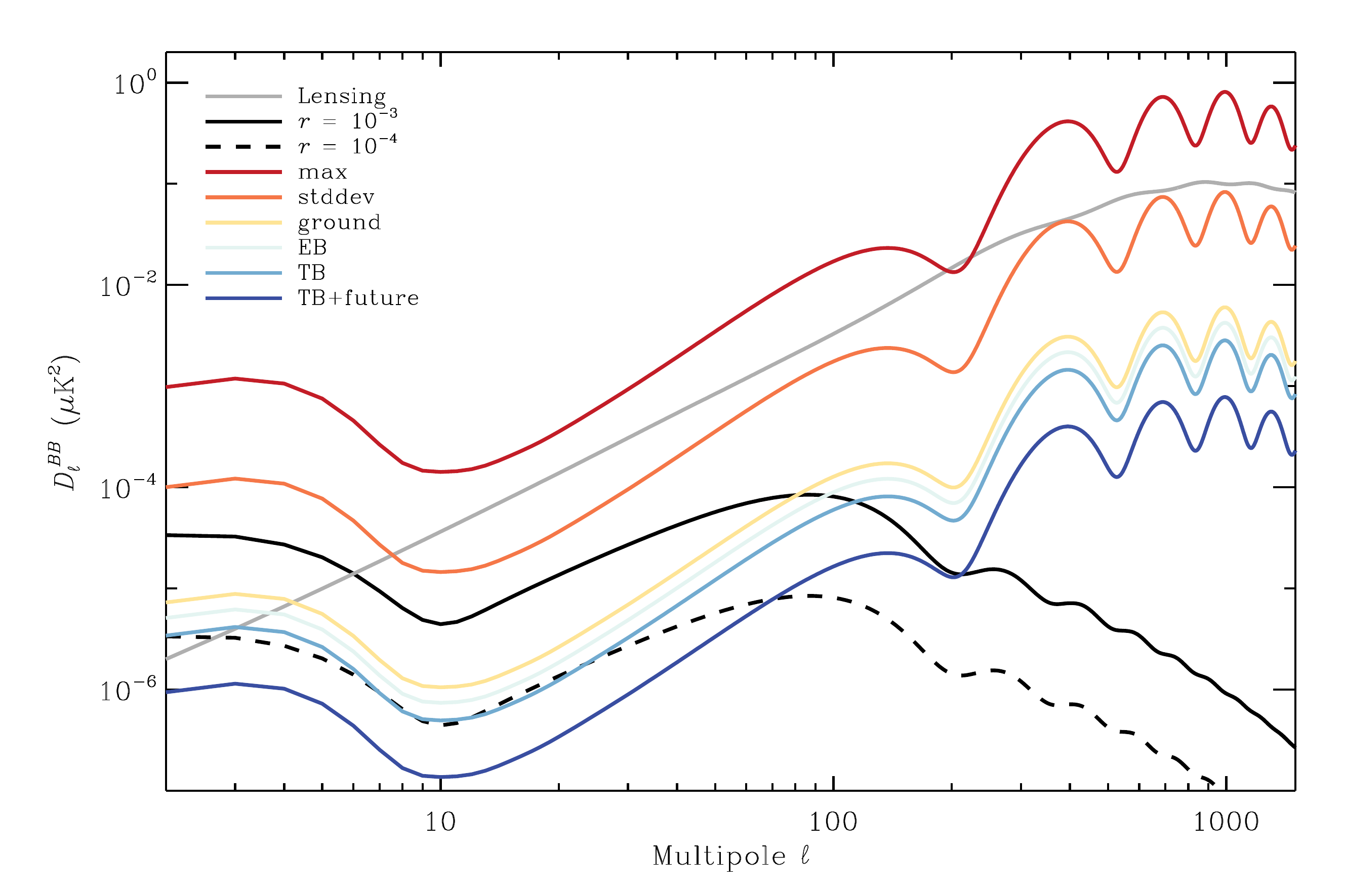}
\includegraphics[height=0.33\textwidth]{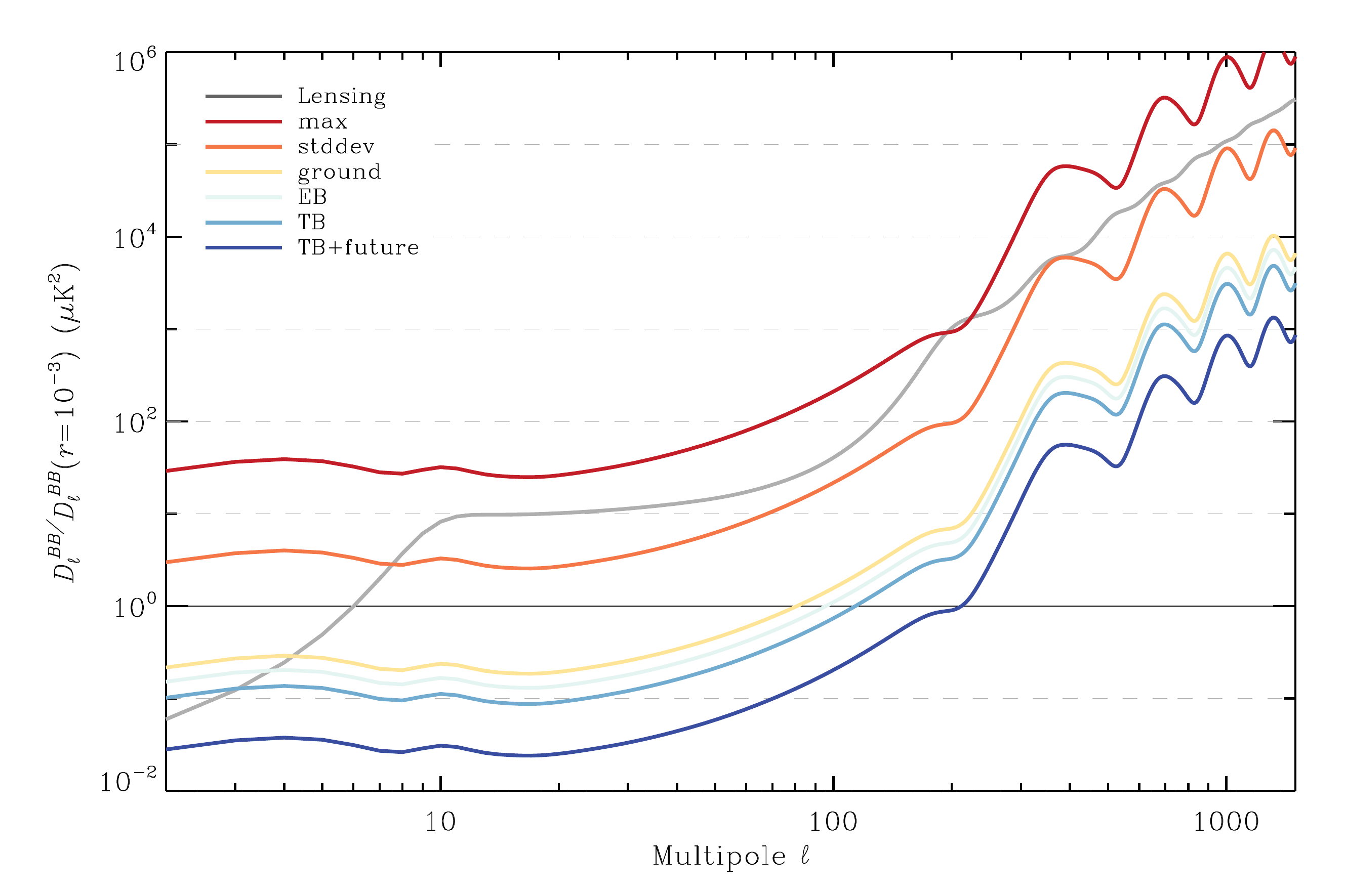}
\caption{\footnotesize {\bf Left panel}: $\Delta D_\ell^{BB}\equiv\ell(\ell+1)/(2\pi)\cdot\Delta C_\ell^{BB}$ power spectrum bias from $E$-$B$ mixing due to the mis-calibration of the absolute polarisation angle. This bias is plotted for the different absolute calibration errors \deltapsigal{} presented in Sect.~\ref{sect:measurements} (from red to blue, see legend). The \citet{planckcosmo} $\Lambda$CDM best fit $D_\ell^{BB}$ primordial tensor model for $r=10^{-3}$ and $r=10^{-4}$ (solid and dashed black lines, respectively) and $D_\ell^{BB}$ lensing model (gray line) are also displayed. {\bf Right panel}: Same as left panel, but relative to the primordial tensor model for $r=10^{-3}$.}
\label{fig:bb_bias}
\end{figure*}

\section{E-B mixing from absolute polarisation angle mis-calibration}
\label{sec:ebmixing}

A mis-calibration of the absolute polarisation angle by $\Delta\psi_{Gal}$ will lead to a mixing of $E$ and $B$ modes. As far as the CMB is concerned and given the fact that $C_\ell^{EE}$ is much larger than $C_\ell^{BB}$, it is often referred to as an ``$E$ to $B$ leakage'' and reads (e.~g.~\citet[][]{rosset}):

\begin{align}
\tilde{C}_\ell^{BB} &= C_\ell^{BB}\cos^2{2\deltapsigal} + C_\ell^{EE}\sin^2{2\deltapsigal}\nonumber\\
\Leftrightarrow\Delta C_\ell^{BB} &\simeq (2\deltapsigal)^2C_\ell^{EE}.
\label{eq:eb-mixing}
\end{align}

\noindent where $\tilde{C}_\ell^{BB}$ is the effectively measured $C_\ell^{BB}$ spectrum and $\Delta C_\ell^{BB}$ is the corresponding spurious bias component. The $E$ to $B$ leakage is therefore constrained by the error on the absolute angle calibration. Unlike some other systematic effects specific to polarisation, it does not depend on the scan pattern of the observation and therefore cannot be mitigated.

If one uses the Crab nebula as a calibrator, the uncertainty on its polarisation angle \deltapsigal{} sets a lower limit on the calibration error, and this has an impact on the magnitude of the corresponding $B$-modes bias. Fig.~\ref{fig:bb_bias} shows the bias $\Delta C_\ell^{BB}$ for the different combinations of experimental uncertainties presented in Sect.~\ref{sect:measurements}. We see that when we relax the assumption of a constant Crab polarisation angle from 23 to 353\,GHz (\texttt{max} and \texttt{stddev}), the spurious $B$-mode signal from $E$-$B$ mixing exceeds the primordial signal for $r=10^{-3}$ at \emph{all the angular scales}. If we assume the Crab polarisation angle to be constant (\texttt{TB+future}, \texttt{TB}, \texttt{EB} and \texttt{ground}), the biases range from $\sim3$ to $\sim30$\,\% of the primordial tensor signal for $\ell<10$, from $\sim20$ to more than $100$\,\% at $\ell\sim100$ and exceeds the signal in all cases for $\ell>250$.


\section{Likelihood analysis}
\label{sec:likelihood}

In order to quantify the $E$-$B$ mixing effect due to the absolute polarisation angle mis-calibration, we perform a likelihood analysis. In each simulation, we consider a $\tilde{C}_\ell^{BB}$ measurement for $r=0$ and $\deltapsigal\neq0$, reading $\tilde{C}_\ell^{BB}=C_\ell^{BB,{\rm lens.}}+\Delta C_\ell^{BB}(\deltapsigal)$. The lensing only $C_\ell^{BB,{\rm lens.}}$ spectrum is computed from the \citet{planckcosmo} $\Lambda$CDM cosmology and the $\Delta C_\ell^{BB}(\deltapsigal)$ $E$-$B$ mixing component comes from Eq.~\ref{eq:eb-mixing}. In each simulation, we draw randomly the \deltapsigal{} mis-calibration from a Gaussian distribution having a 1\,$\sigma$ dispersion corresponding to the error in each of the cases presented in Sect.~\ref{sec:combined}. The log-likelihood $\log\left(\mathcal{L}(r)\right)=\chi^2(r)/2$ then reads:

\begin{align}
\label{eq:likelihood}
&2\log\left(\mathcal{L}(r)\right)=\chi^2(r)\nonumber\\ 
&=\sum_{\ell\in[\ell_{\rm min},\ell_{\rm max}]}\frac{\left(\tilde{C}_\ell^{BB}-r\cdot C_\ell^{BB,r=1}-C_\ell^{BB,{\rm lens.}}\right)^2}{\sigma_{\rm tot.}^2}\nonumber\\
&=\sum_{\ell\in[\ell_{\rm min},\ell_{\rm max}]}\frac{\left(\Delta C_\ell^{BB}(\deltapsigal)-r\cdot C_\ell^{BB,r=1}\right)^2}{\sigma_{\rm tot.}^2},
\end{align}


\noindent where $C_\ell^{BB,r=1}$ is the \citet{planckcosmo} $\Lambda$CDM cosmology tensor mode spectrum for $r=1$ and $\sigma_{\rm tot}$ the quadratic sum of the cosmic variance and the 1\,$\sigma$ $E$-$B$ mixing residual term. The cosmic variance is computed for $f_{\rm sky}=0.5$, assuming a 10\,\% residual after delensing.

\begin{figure*}
\centering
\includegraphics[height=0.4\textwidth]{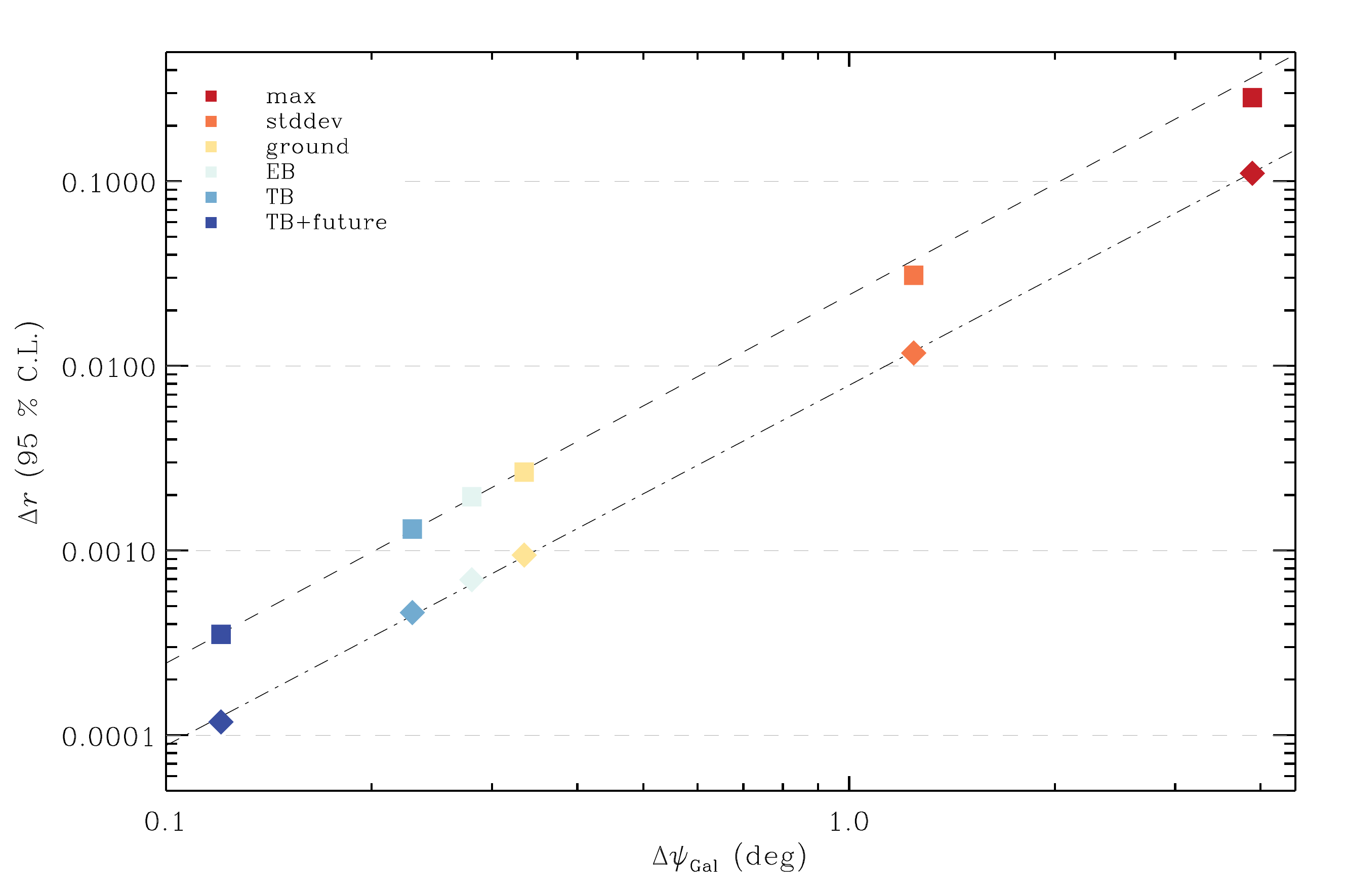}
\caption{\footnotesize Likelihood posterior on $r$ biases (with respect to an input signal of $r=0$) for the different cases of combined calibration errors (presented in Sect.~\ref{sect:measurements}) from 10\,000 Monte-Carlo simulations, as a function of the combined error on the angle $\Delta\psi_{\rm Gal}$ in degrees. They are computed independently for the recombination ($30<\ell<300$, squares) and the reionisation ($2<\ell<30$, diamonds) bumps. The best-fit $\Delta r=A\Delta\psi_{\rm Gal}^\beta$ power-laws are displayed as dashed and dashed-dotted solid black lines.
\label{fig:posteriors}}
\end{figure*}

The likelihood function is computed on 10\,000 Monte-Carlo simulations. For each simulation, we build the posterior on $r$ from Eq.~\ref{eq:likelihood} and fit the bias $\Delta r$ with respect to $r=0$. The 10\,000 biases follow a typical $\chi^2$ distribution. We sort these $\Delta r$ biases and find the value $\Delta r(\rm 95\,\%\,C.L.)$ defined as the $r$ value for which 95\,\% of the simulations have a smaller $\Delta r$. This is done for the recombination bump ($\ell_{\rm min}=30$, $\ell_{\rm max}=300$) and the reionisation bump ($\ell_{\rm min}=2$, $\ell_{\rm max}=30$).

Neither foregrounds nor their residuals are modelled in this simple analysis, in order to focus on the effect of the polarisation angle mis-calibration. Thus, in addition to assume a perfect component separation, we suppose that the mis-calibration $E$-$B$ mixing residual from foregrounds is as well perfectly removed. This is a good approximation at first order, as the $E$-$B$ mixing term doesn't change their spectral colors.

The $\Delta r(\rm 95\,\%\,C.L.)$ values are presented in Fig.~\ref{fig:posteriors} for the recombination and reionisation bumps. We can see that the spurious $B$-mode polarisation coming from $E$-$B$ mixing is more penalising at high-$\ell$, resulting in higher $r$ biases for the recombination bump than for the reionisation bump. The two cases considered in Sect.~\ref{sec:combined} where we do not assume a spectrally constant polarisation for the Crab nebula (\texttt{max} and \texttt{stddev}) lead to biases on the $r$ posterior that are of the order of $r=10^{-2}$ or larger. In the cases where we assume that the Crab polarisation angle is constant (\texttt{ground}, \texttt{EB}, \texttt{TB} and \texttt{TB+future}), the biases on $r$ range from $r\sim10^{-4}$ to $r\sim3\times10^{-3}$. For the detection of $r=10^{-2}$, the best \emph{current} combined uncertainty on the Crab polarisation angle (\texttt{TB} case) would lead to a potential 95\,\% C.L. bias of $\sim10$\,\% at the recombination bump and $\sim4$\,\% at the reionisation bump. With respect to $r=10^{-3}$, the current limits would lead to a 100\,\% bias at the recombination bump and 40\,\% at the lowest $\ell$ multipoles. Considering new measurements of the Crab polarisation angle, as in the \texttt{TB+future} case, the bias could be shrunk down to negligible values for the measurement of $r=10^{-2}$ and down to $\sim10$ and $\sim30$\,\% of $r=10^{-3}$, for the reionisation and recombination bumps respectively.

From Eq.~\ref{eq:eb-mixing}, we expect that the bias on $r$ due to $E$-$B$ mixing from an incorrect calibration of the absolute polarisation angle would scale as $\Delta r(\rm 95\,\%\,C.L.)\propto\Delta\psi_{\rm Gal.}^2$. We have fitted the biases on $r$ from our likelihood analysis by power-laws of the form ${\Delta r(\rm 95\,\%\,C.L.)=A\cdot(\Delta\psi_{\rm Gal.})^\beta}$ (see Fig.~\ref{fig:posteriors}). We find for the reionisation bump $(A,\beta)_{2<\ell<30}=(0.008,1.95)$ and for the recombination bump, $(A,\beta)_{30<\ell<300}=(0.024,2.00)$.


\section{Conclusion and discussion}
\label{sec:conclusion}

In the present work, we study a compendium of the best constraints on the Crab nebula polarisation angle to date, from 23 to 353\,GHz \citep{crabwmap,crabxpol,crablfi,crabnika} to derive the combined uncertainty on this angle under different assumptions. We explore the effect such an uncertainty has on the measurement of the CMB $B$-mode primordial signal, through the bias it generates on the estimation of the $r$ parameter, if one uses the Crab nebula as a calibrator for the absolute polarisation angle of an experiment. No other source of $r$ biases is considered in this work. 

We find that, in order to prevent biases larger than $r=10^{-2}$, one must assume that the Crab polarisation angle is constant across microwave frequencies. This is a fair hypothesis, given that current studies including \citet{crabnika} are compatible with a single synchrotron component being responsible for the Crab nebula microwave emission. Nevertheless, the current measurement systematic errors and dispersion are large and future constraints might be needed to strengthen these constraints.

If we assume the Crab polarisation angle as constant from 23 to 353\,GHz, and consider the ground calibration errors for the \expe{Planck-HFI} measurements, the combined uncertainty on $\psi_{Gal}$ leads to potential biases on $r$ of the order of $3\times10^{-3}$ at the recombination bump and $\sim10^{-3}$ at the reionisation bump. Our estimates address absolute polarisation angle calibration uncertainty. The consequent biases would thus show for any experiment, independently of its sensitivity, and they jeopardize the measure primordial CMB $B$-modes around $r=10^{-3}$, as currently targeted by ongoing and near future projects.

The \expe{Planck-HFI} uncertainty on the Crab polarisation angle measurements can be narrowed by considering the errors coming from the $C_\ell^{EB}$ and $C_\ell^{TB}$ minimisations. In the latter case, the $r$ bias arising from the incorrect calibration of the absolute polarisation angle is $\sim4\times10^{-4}$ at the recombination bump and $\sim10^{-4}$ at the reionisation bump. However, these minimisations make the assumption that the \expe{Planck-HFI} $C_\ell^{EB}$ and $C_\ell^{TB}$ are not contaminated by Galactic components or systematic effects beyond the calibration of the instrumental absolute polarisation angle. 

The present study suggests that the error on $r$ coming from the absolute polarisation angle calibration would be mitigated when adding extra measurements of the Crab polarisation angle. We find that if we add two future measurements with total uncertainties of $0.2\,^\circ$ to the current observations, the bias on $r$ from mis-calibration goes down to $\sim4\times10^{-4}$ at the recombination bump and $\sim10^{-4}$ at the reionisation bump. These values are acceptable for an experiment targeting $r=10^{-3}$, especially at large angular scales. However, these new measurements will not only be needed to reduce the statistical uncertainty on the Crab nebula polarisation angle. They are required to definitively assess its stability across the microwave frequency. The \expe{Xpol} \citep{xpol} and \expe{Nika2} \citep{nikaII} instruments could allow to achieve such measurements at 90 and 260\,GHz.

In this paper, we combine measurements of the Crab nebula polarisation angle from experiments observing with a wide range of angular resolutions. By directly comparing these measurements, we assume that the aperture photometry (or similar techniques) captures all the emission from the Crab, and that they are not contaminated by other sources of emission. Naturally, an additional complication in using the Crab nebula as an absolute polarisation angle calibrator for any given CMB experiment will come from the uncertainties in the knowledge of the instrumental polarised beams. The effect of an incorrect beam modeling, including sidelobes, requires a case-by-case analysis and goes beyond the scope of this paper.

The polarisation efficiency is another crucial instrumental parameter that has to be characterised by an experiment aiming at the measurement of the CMB $B$-modes. The Crab polarised intensity could be used as a calibrator for this parameter. Nevertheless, unlike the polarisation angle, the Crab polarised intensity is not constant across frequencies \citep{crabnika}. Therefore the expected final polarisation efficiency calibration uncertainty is limited by frequency extrapolation of the Crab nebula emission. Moreover, the uncertainty on the annual fading of the Crab synchrotron emission will affect the calibration of the polarisation efficiencies, while it is not expected to influence the determination of the polarisation angle. 


\begin{acknowledgements}
We thank the \expe{Planck} Collaboration for allowing us to use the 2018 \expe{Planck} maps in advance of public release to obtain integrated flux densities in intensity and polarisation in the Crab nebula. We thank Douglas Scott for useful comments on the paper.
\end{acknowledgements}


\bibliographystyle{aat}

\bibliography{crab_calibration}


\end{document}

%% file: Planck.tex
\def\setsymbol#1#2{\expandafter\def\csname #1\endcsname{#2}}
\def\getsymbol#1{\csname #1\endcsname}

\newbox\tablebox    \newdimen\tablewidth
\def\leaderfil{\leaders\hbox to 5pt{\hss.\hss}\hfil}
\def\tablenote#1 #2\par{\begingroup \parindent=0.8em
    \abovedisplayshortskip=0pt\belowdisplayshortskip=0pt
    \noindent
    $$\hss\vbox{\hsize\tablewidth \hangindent=\parindent \hangafter=1 \noindent
    \hbox to \parindent{$^#1$\hss}\strut#2\strut\par}\hss$$
    \endgroup}

\def\L2{\ifmmode L_2\else $L_2$\fi}

\def\DeltaT{\ifmmode \Delta T\else $\Delta T$\fi}
\def\deltat{\ifmmode \Delta t\else $\Delta t$\fi}
\def\fknee{\ifmmode f_{\rm knee}\else $f_{\rm knee}$\fi}
\def\Fmax{\ifmmode F_{\rm max}\else $F_{\rm max}$\fi}
\def\solar{\ifmmode{\rm M}_{\mathord\odot}\else${\rm M}_{\mathord\odot}$\fi}
\def\Msolar{\ifmmode{\rm M}_{\mathord\odot}\else${\rm M}_{\mathord\odot}$\fi}
\def\Lsolar{\ifmmode{\rm L}_{\mathord\odot}\else${\rm L}_{\mathord\odot}$\fi}
\def\inv{\ifmmode^{-1}\else$^{-1}$\fi}
\def\mo{\ifmmode^{-1}\else$^{-1}$\fi}
\def\sup#1{\ifmmode ^{\rm #1}\else $^{\rm #1}$\fi}
\def\expo#1{\ifmmode \times 10^{#1}\else $\times 10^{#1}$\fi}
\def\,{\thinspace}
\def\lsim{\mathrel{\raise .4ex\hbox{\rlap{$<$}\lower 1.2ex\hbox{$\sim$}}}}
\def\gsim{\mathrel{\raise .4ex\hbox{\rlap{$>$}\lower 1.2ex\hbox{$\sim$}}}}

\def\simprop{\mathrel{\raise .4ex\hbox{\rlap{$\propto$}\lower 1.2ex\hbox{$\sim$}}}}
\def\deg{\ifmmode^\circ\else$^\circ$\fi}
\def\pdeg{\ifmmode $\setbox0=\hbox{$^{\circ}$}\rlap{\hskip.11\wd0 .}$^{\circ}
          \else \setbox0=\hbox{$^{\circ}$}\rlap{\hskip.11\wd0 .}$^{\circ}$\fi}
\def\arcs{\ifmmode {^{\scriptstyle\prime\prime}}
          \else $^{\scriptstyle\prime\prime}$\fi}
\def\arcm{\ifmmode {^{\scriptstyle\prime}}
          \else $^{\scriptstyle\prime}$\fi}
\newdimen\sa  \newdimen\sb
\def\parcs{\sa=.07em \sb=.03em
     \ifmmode \hbox{\rlap{.}}^{\scriptstyle\prime\kern -\sb\prime}\hbox{\kern -\sa}
     \else \rlap{.}$^{\scriptstyle\prime\kern -\sb\prime}$\kern -\sa\fi}
\def\parcm{\sa=.08em \sb=.03em
     \ifmmode \hbox{\rlap{.}\kern\sa}^{\scriptstyle\prime}\hbox{\kern-\sb}
     \else \rlap{.}\kern\sa$^{\scriptstyle\prime}$\kern-\sb\fi}
\def\ra[#1 #2 #3.#4]{#1\sup{h}#2\sup{m}#3\sup{s}\llap.#4}
\def\dec[#1 #2 #3.#4]{#1\deg#2\arcm#3\arcs\llap.#4}
\def\deco[#1 #2 #3]{#1\deg#2\arcm#3\arcs}
\def\rra[#1 #2]{#1\sup{h}#2\sup{m}}

\def\dots{\relax\ifmmode \ldots\else $\ldots$\fi}
\def\WHzsr{\ifmmode $W\,Hz\mo\,sr\mo$\else W\,Hz\mo\,sr\mo\fi}
\def\mHz{\ifmmode $\,mHz$\else \,mHz\fi}
\def\GHz{\ifmmode $\,GHz$\else \,GHz\fi}
\def\mKs{\ifmmode $\,mK\,s$^{1/2}\else \,mK\,s$^{1/2}$\fi}
\def\muKs{\ifmmode \,\mu$K\,s$^{1/2}\else \,$\mu$K\,s$^{1/2}$\fi}
\def\muKRJs{\ifmmode \,\mu$K$_{\rm RJ}$\,s$^{1/2}\else \,$\mu$K$_{\rm RJ}$\,s$^{1/2}$\fi}
\def\muKHz{\ifmmode \,\mu$K\,Hz$^{-1/2}\else \,$\mu$K\,Hz$^{-1/2}$\fi}
\def\MJysr{\ifmmode \,$MJy\,sr\mo$\else \,MJy\,sr\mo\fi}
\def\MJysrmK{\ifmmode \,$MJy\,sr\mo$\,mK$_{\rm CMB}\mo\else \,MJy\,sr\mo\,mK$_{\rm CMB}\mo$\fi}
\def\microns{\ifmmode \,\mu$m$\else \,$\mu$m\fi}

\def\muK{\ifmmode \,\mu$K$\else \,$\mu$\hbox{K}\fi}
\def\microK{\ifmmode \,\mu$K$\else \,$\mu$\hbox{K}\fi}
\def\muW{\ifmmode \,\mu$W$\else \,$\mu$\hbox{W}\fi}
\def\kms{\ifmmode $\,km\,s$^{-1}\else \,km\,s$^{-1}$\fi}
\def\kmsMpc{\ifmmode $\,\kms\,Mpc\mo$\else \,\kms\,Mpc\mo\fi}
\providecommand{\sorthelp}[1]{}